\documentclass[review]{elsarticle}

\usepackage{lineno,hyperref}
\usepackage[utf8x]{inputenc} 
\usepackage{graphicx}
\usepackage{color}
\usepackage[boxruled,vlined,linesnumbered] {algorithm2e}
\usepackage{multirow}
\usepackage{amssymb}
\usepackage[table]{xcolor}
\usepackage{float}
\usepackage{hyperref}
\hypersetup{
    colorlinks=true,
    linkcolor=blue,
    filecolor=magenta,      
    urlcolor=cyan,
}

\usepackage[boxruled,vlined,linesnumbered]{algorithm2e}

\usepackage[a4paper, total={6in, 10in}]{geometry}
\modulolinenumbers[5]

\SetKwRepeat{Do}{do}{while}%

\journal{IET Book Chapter}

\begin{document}

\begin{frontmatter}


\title{Fuzzy Adaptive Tuning of a Particle Swarm Optimization Algorithm for Variable-Strength Combinatorial Test Suite Generation}

\author{Kamal Z. Zamli}
\address{Faculty of Computer Systems and Software Engineering, University Malaysia Pahang, Gambang, Malaysia

kamalz@ump.edu.my}

\author{Bestoun S. Ahmed*}

\address{Department of Computer Science, Faculty of Electrical Engineering, Czech Technical University, Karlovo n´am. 13, 121 35 Praha 2, Czech Republic

albeybes@fel.cvut.cz}

\author{Thair Mahmoud}
\address{School of Engineering, Edith Cowan University, 270 Joondalup Dr, Joondalup, WA6027, Australia, 

t.mahmoud@ecu.edu.au}

\author{Wasif Afzal}
\address{School of Innovation, Design and Engineering, Mälardalen University, Sweden

wasif.afzal@mdh.se}

\begin{abstract}
\linespread{1.1}\selectfont

Combinatorial interaction testing is an important software testing technique that has seen lots of recent interest. It can reduce the number of test cases needed by considering interactions between combinations of input parameters. Empirical evidence shows that it effectively detects faults, in particular, for highly configurable software systems. In real-world software testing, the input variables may vary in how strongly they interact; variable strength combinatorial interaction testing (VS-CIT) can exploit this for higher effectiveness. 
The generation of variable strength test suites is a non-deterministic polynomial-time (NP) hard computational problem \cite{BestounKamalFuzzy2017}. Research has shown that stochastic population-based algorithms such as particle swarm optimization (PSO) can be efficient compared to alternatives for VS-CIT problems. Nevertheless, they require detailed control for the exploitation and exploration trade-off to avoid premature convergence (i.e. being trapped in local optima) as well as to enhance the solution diversity. Here, we present a new variant of PSO based on Mamdani fuzzy inference system \cite{Camastra2015,TSAKIRIDIS2017257,KHOSRAVANIAN2016280}, to permit adaptive selection of its global and local search operations. We detail the design of this combined algorithm and evaluate it through experiments on multiple synthetic and benchmark problems. 

We conclude that fuzzy adaptive selection of global and local search operations is, at least, feasible as it performs only second-best to a discrete variant of PSO, called DPSO. Concerning obtaining the best mean test suite size, the fuzzy adaptation even outperforms DPSO occasionally. We discuss the reasons behind this performance and outline relevant areas of future work.

\end{abstract}

%\begin{keyword}
%Constrained combinatorial interaction \sep Multi-objective particle swarm optimisation \sep Test generation tools \sep Search-based software engineering \sep Test case design techniques

%\end{keyword}

%\linenumbers

\end{frontmatter}

\section{Introduction}

Interaction of input parameters for software systems has been introduced as one of the primary sources of failure. Traditional test design techniques are useful for fault discovery and prevention when a single input value causes the fault at a time. However, they may not be adequate to handle faults caused by the interaction of several inputs. Combinatorial Interaction Testing (CIT) (sometimes called $t$-way testing where $t$ is the combination strength) addressed this problem properly by taking the interaction of two or more inputs and generating a combinatorial test suite that helps in early fault detection in the testing life cycle \cite{Nie2011, Yilmaz2014}. The key insight of this testing approach is that, not every input parameter of the system contribute to the faults of the system and most of the faults are addressed by including only a few number of the interactions of input parameters \cite{Nie2011}. Thus, CIT can be a useful approach to minimize the number of test cases needed and thus be a practical way forward given that exhaustive testing is almost never an option for modern, complex software systems.

For instance, pairwise testing (i.e., $2$-way testing) is a common approach and has been applied effectively in many practical testing situations \cite{Hervieu2016}. However, empirical evidence has shown that software failures may be triggered by unusual combinations involving more than two input parameters and values \cite{Nie2011,Kacker2013,Kuhn2013}. Such cases have been identified as $t$-way CIT where $t>2$ and there is an equal (or uniform) interaction between the values. But in many cases, the evidence also show that the interaction strength cannot always be considered the same or uniform \cite{Ahmed2011,ZamliISPaper2017,1245373}. Rather the interaction may vary between groups of inputs. This has been described as a particular type of CIT and called the variable-strength CIT (VS-CIT) \cite{huang2013prioritizing}.

As in the case of CIT, the VS-CIT is a non-deterministic polynomial-time (NP) hard computational problem. In the case of CIT, strategies based on heuristic methods and artificial intelligent (AI) have proved to be efficient to generate smaller test suites for a certain level of interaction coverage. Although many strategies have been developed based on such methods, few of the published methods support the generation of variable-strength (VS-CIT) test suites. For instance, Simulated Annealing (SA) is one of the most successful strategies to generate VS-CIT test suites. More recently, we have developed a more flexible and practical strategy to generate VS-CIT test suites based on Particle Swarm Optimization (PSO) \cite{Ahmed2011,Ahmed2012,BestounConstraints2017,AhmedBestoun2016}. 

Although showing efficient results as compared to other strategies, our experience and investigations of the PSO literature indicate that the quality of the results produced by the PSO search algorithm is mainly dependent on the search adaptation parameters. 

The rationale behind this situation is that the search mechanism of PSO is, in fact, a combination of exploration and exploitation. 

The exploration is needed to find a globally optimal solution while the exploitation is required to achieve local refinement of solutions towards more promising candidates. To get quality solutions, a careful tuning must typically be made of the search parameters to trade-off effectively between (global) exploration and (local) exploitation. Evidence have shown that the values for these parameters may vary depending on the problem complexity itself as well as other characteristics of the search \cite{Wang2014,Xu2013}.  

In previous research \cite{Ahmed2012,Bestoun2012IJICIC}, we have tuned these parameters experimentally to find the best values. Although useful, these best values obtained for one configuration and problem may not be the best for other configurations or problems. Thus we have recently developed an adaptive mechanism to tune these parameters for each configuration \cite{Mahmoud2015}. However, for the VS-CIT problem, the complexity of the search increases even more and adaptations might be needed many times during the generation of a test suite. Hence, there is a need for an even more dynamic and adaptive approach to tune the search parameters. In this chapter, we extend our previous works on CIT and VS-CIT using an adaptive and dynamic approach for tuning the searching parameters of PSO using Mamdani fuzzy inference system (FIS). The strategy can monitor the performance of the PSO and adjust the search process during its execution to improve its efficiency. 

The rest of this chapter is organized as follows. Section \ref{CITSection} gives a brief introduction to CIT including the mathematical objects used with it and a motivating real-world example to show how the VS-CIT could be used with practical testing. Section \ref{RelatedWork} gives an extensive review of the related work in the context of VS-CIT. Then Section \ref{FuzzyAdaptive VS-CIT} discusses the essential components of our approach for fuzzy adaptive VS-CIT. Section \ref{Evaluation} illustrates the evaluation process of our approach. Sections \ref{Evaluation} and \ref{Discussion} discusses the results and the observation of the evaluation process. Finally, section \ref{Conclusion} gives the concluding remarks.

\section{Combinatorial Interaction Testing} \label{CITSection}

\subsection{Preliminaries}

Theoretically, the combinatorial test suite depends on a well-known mathematical object called Covering Array (CA). Originally, CA has been gained more attention as a practical alternative of oldest mathematical object called Orthogonal Array (OA) that has been used for statistical experiments \cite{Cohen2003}.

An $OA_\lambda (N; t, k, v)$ is an $N \times k$ array, where for every $N \times t$ sub-array, each $t-tuple$ occurs exactly $\lambda$ times, where $\lambda = N/v^t$; $t$ is the combination strength; $k$ is the number of input functions ($k \geqslant  t$); and $v$ is the number of levels associated with each input parameter of the software-under-test (SUT)\cite{Colbourn2010,Sloane1993}. Practically, it is very hard to translate these firm rules except for small systems with a small number of input parameters and values. Hence, there is no significant benefit in case of medium and large size SUTs, as it is very hard to generate OA for them. In addition, based on the rules mentioned above, it is not possible to represent OA when there are different levels for each input parameters \cite{Kari2004}.

To address the limitations of OA, CA has been introduced. A $CA\lambda(N ; t, k, v)$ is a $N \times k$ array over $(0, . . . , v - 1)$ such that every $B=\lbrace b_0, ..., b_{t-1} \rbrace $ is $\lambda$-covered and every $N \times t$ sub-array contains all ordered subsets from $v$ values of size $t$ at least $\lambda$ times, where the set of column $B=\lbrace b_0, ..., b_{t-1} \rbrace \supseteq \lbrace0, ..., k-1 \rbrace$ \cite{Ahmed2015,Hartman2005}. In this case, each tuple is to appear at least once in a CA. 

In the case when the number of component values varies, this can be handled by Mixed Covering Array (MCA). A $MCA (N; t, k, (v_1, v_2, … v_k))$, is an $N \times k$ array on $v$ values, where the rows of each $N \times t$ sub-array cover and all $t$ interactions of values from the $t$ columns occur at least once. For more flexibility in the notation, the array can be presented by $MCA (N; t, v_1^{k_1} v_2^{k_2} .. v_k^k)$.

In real world complex systems, the interaction strength may vary between the system parameters. In fact, the interaction of some parameters may be stronger than other parameters, or they may not interact at all. Variable strength covering array (VSCA) is introduced to cater for this issue. A $VSCA (N; t; k, v, (CA_1… CA_k))$ represents $N \times p$ MCA of strength $t$ containing vectors of $CA_1$ to $CA_k$, and a subset of the $k$ columns each of strength $ > t$.

\subsection{Motivating Example} \label{MotivationExampleSection}

\begin{figure}
\centering
\includegraphics [width=5 in]{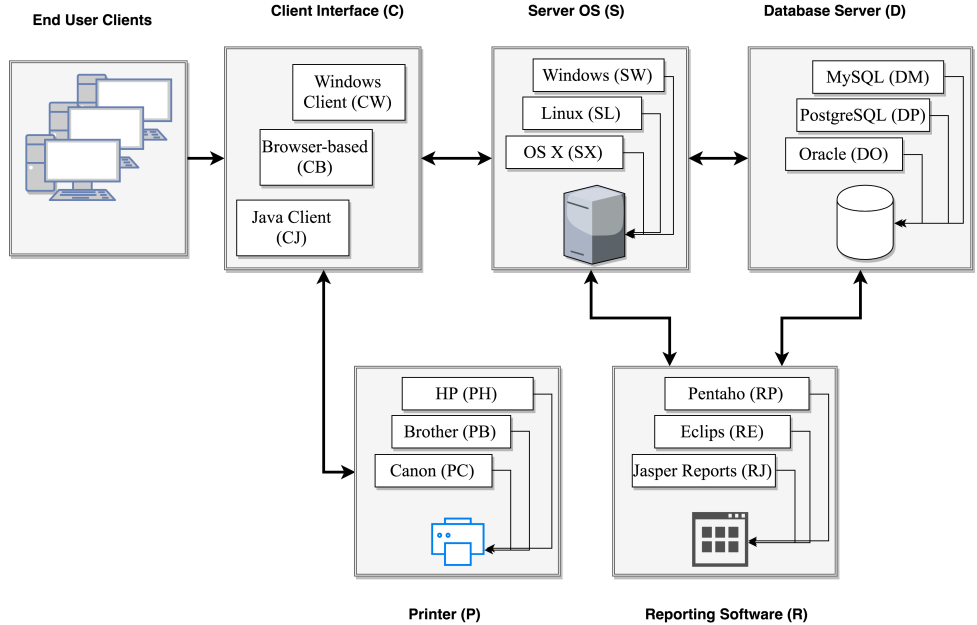}

\caption{Software Architecture of a 3-Tier System Under Test}
\label{MotivatingExample}
\end{figure}

The problem of CA and its variants can be expressed and illustrated in practice through different case studies. Here, we adopt a real case study described by Raaphorst \cite{Raaphorst2013} . The case study illustrates the use of CA for system testing. Figure \ref{MotivatingExample} shows the architecture of a 3-tier system under test. The system may have different components or parameters. Similar to its use in the CA notations, in this chapter the term component ``$k$'' is used to refer the components of the SUT. In addition, the term variable ``$v$'' is used to refer the values or configurations of each component. In the current example, the SUT consist of five components each of which having three values.

The client may communicate with the servers through some interfaces for interacting with the databases. Here, we specify three kinds of interfaces, ``Windows Client (CW)'', ``Browser-based (CB)'' and ``Java Client (CJ)''. The Operating system of the servers may take different kinds like ``Windows (SW)'', ``Linux (SL)'' or ``OS X (SX)''. The database servers also may take different kinds like ``MySQL (DM)''9, ``PostgreSQL (DP)'' or ``Oracle (DO)''. Furthermore, different reporting packages convey the data for users like ``Pentaho (RP)'', ``Eclipse BIRT (RE)'', or ``Jasper Reports (RJ)''. The user then may print these reports on different printer machines like ``HP (PH)'', ``Brother (PB)'', or ``Canon (PC)'' through the interface.

To test the system through its different configurations, there are $3^5 = 243$ configurations to be tested in case of an exhaustive testing process. Such testing process will be too costly with the growing of the system under test when the architecture of the system becomes more complicated. These test cases could be minimized effectively using the CIT through CA and VSCA mathematical objects.

By following the CIT process for the system components, we can get two significant benefits. First, we can optimize the test cases dramatically as compared to the exhaustive test suite. Second, we can test the system to find the faults that may happen due to the inappropriate combinations in the configurations. Here, we can test the system against different interaction strengths. For example, Table \ref{CITTableMotivationExample} depicts the test suite of the system when the interaction strength $t=2$. Here, all the interactions of two components are covered by only 15 test cases, where \textbf{T} is the test number in the table.

\begin{table}
\centering
\caption{CIT test suite for the System Architecture in Figure \ref{MotivatingExample}}

\includegraphics [width=4.1 in]{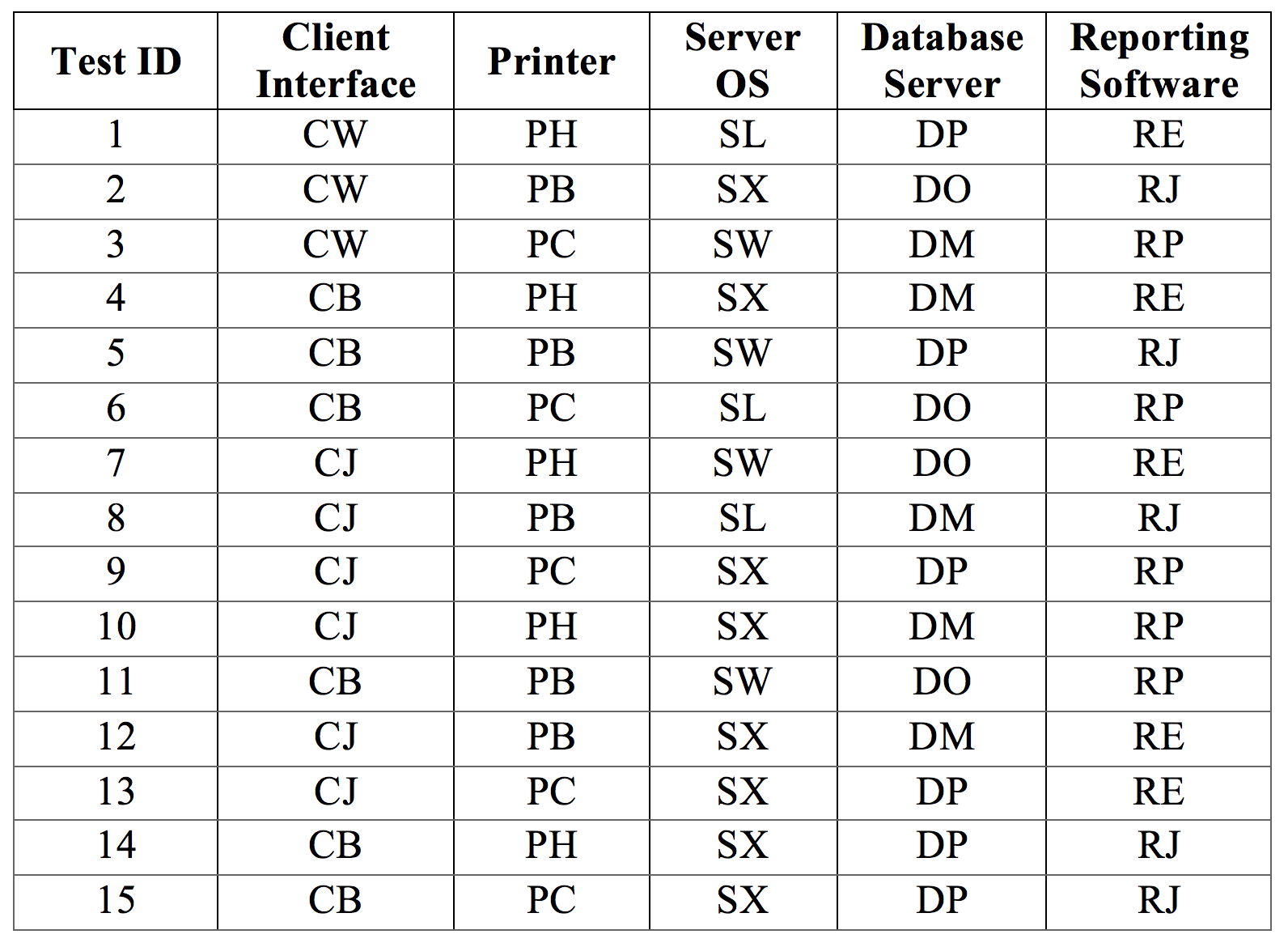}

\label{CITTableMotivationExample}

\end{table}

\begin{figure}
\centering
\includegraphics [width=2.5 in]{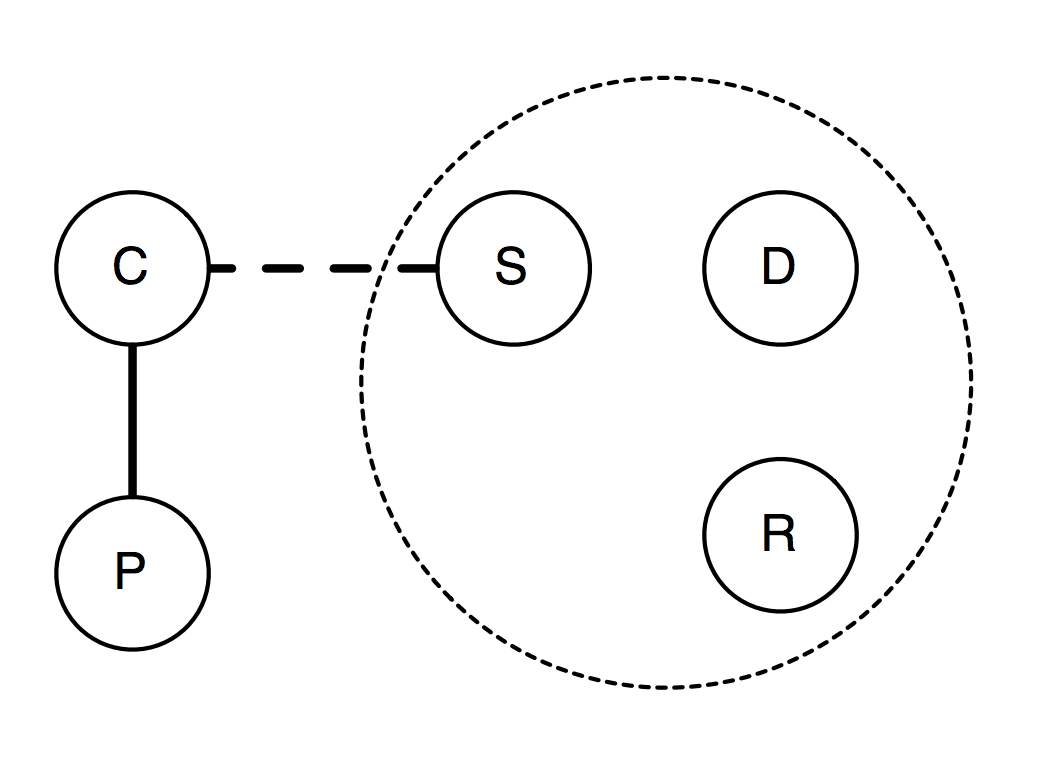}

\caption{Hypergraph to represent VS-CIT of the Components in Figure \ref{MotivatingExample} \cite{Raaphorst2013}}
\label{hypergraphExample}
\end{figure}

Sometimes, the fault may happen due to the interaction of sub-configurations. For example, we may take the interaction of three components for the whole system, but the interaction of two components for a sub-set of the system components also. Hence, we can identify the components that have direct interaction with each other. Figure \ref{hypergraphExample} shows a hypergraph that model this kind of interaction apparently.

As can be seen clearly in Figure \ref{hypergraphExample}, the whole system may have specific interaction strength. However, the ``Server OS (S)'', ``Database Server (D)'' and ``Reporting Software (R)'' has direct interaction with each other and they may have different interactions among them. Here, VSCA can address this issue effectively. Table \ref{VS-CITTableMotivationExample} shows this test suite following the VSCA concepts taking interaction strength $t=3$ for the whole system and $t=2$ among the the ``Server OS (S)'', ``Database Server (D)'' and ``Reporting Software (R)'' components. Here, we can cover this configuration of the system by only 27 test cases instead of the exhaustive test suite. 

\begin{table}
\centering
\caption{The VS-CIT Test Suite for the Architecture System in Figure \ref{MotivatingExample}}

\includegraphics [width= 3 in]{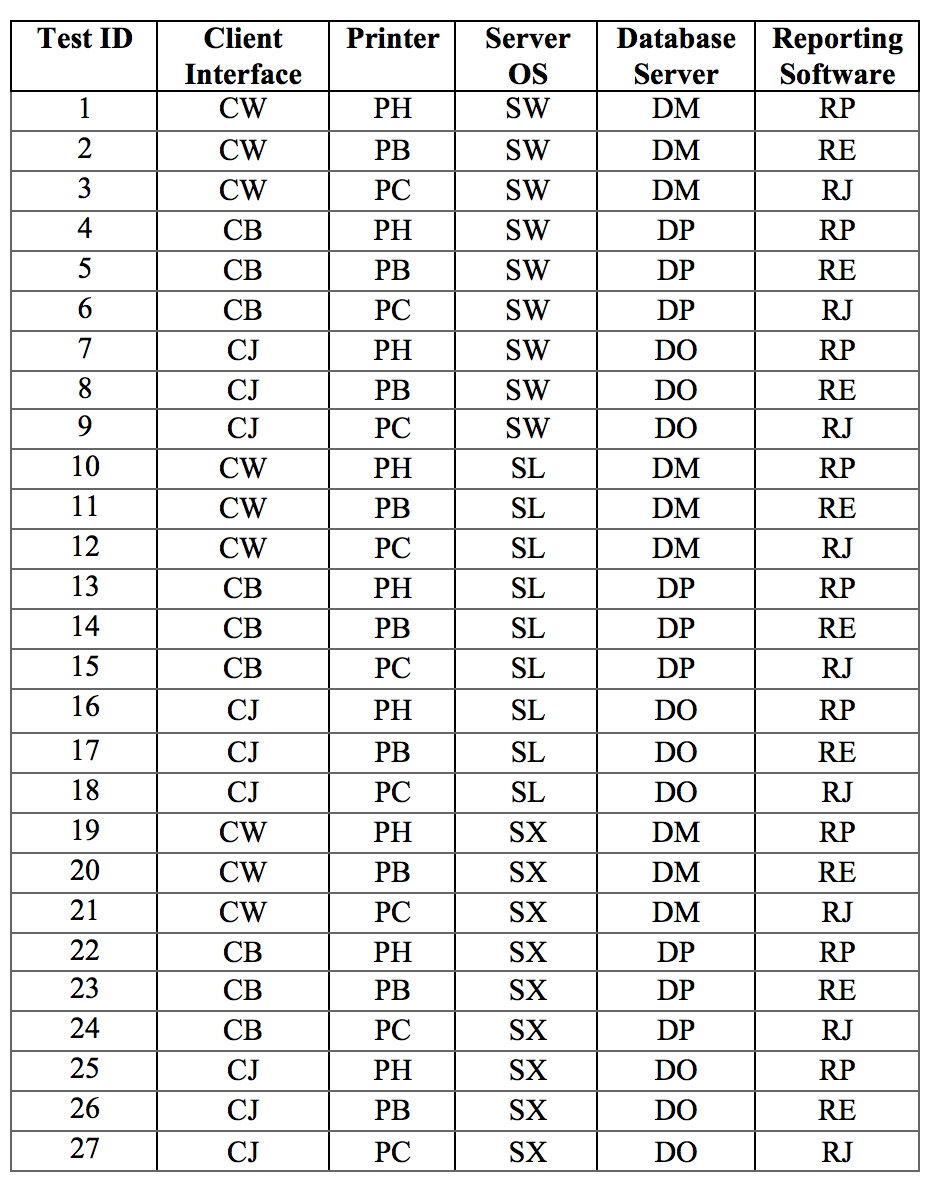}

\label{VS-CITTableMotivationExample}

\end{table}

\section{Related Work} \label{RelatedWork}

As discussed previously, VS-CIT is about creating the combinatorial mathematical object VSCA. VSCA is practical in cases where it is desirable to test a subset of components involved in high-degree interactions while other components may only require testing at low interaction strength. While a plethora of techniques exists to construct CA and MCA (see, e.g.,~\cite{Bestoun-plos,Nie2011} for a brief overview), relatively fewer papers focus on optimal VSCA construction. VSCA, as with CA construction in general~\cite{Nie2011}, is an NP-hard problem. Subsequently, greedy and metaheuristic approaches have been applied for VSCA construction.

Cohen at al.~\cite{1245373} used ten runs of a simulated annealing (SA) algorithm to get minimum, maximum and average sizes of VSCAs. The use of SA might not be the most efficient way for VSCA construction since the process involves a binary search through a large random array. However, SA does generate VSCAs of known minimum sizes. Ziun \textit{et al.}~\cite{4601539} used a greedy heuristic strategy to improve the efficiency of VSCA construction. They proposed algorithms based on one-test-at-a-time \cite{Bryce2007} and in-parameter-order (IPO) strategy respectively \cite{Forbes2008,Lei2008}. Chen \textit{et. al.}~\cite{Chen2009} used a combination of ant colony system (ACS) and one-test-at-a-time strategy to build variable strength test suites. Finally, Bestoun and Zamli~\cite{Ahmed2011} implemented a particle swarm optimization algorithm to generate VS-CIT test suites and reported improved performance for high interaction strengths. Some tools also support VS-CIT, e.g., Test Vector Generator (TVG)\footnote{https://sourceforge.net/projects/tvg/} and Pairwise Independent Combinatorial Testing (PICT)~\cite{Czerwonka2008}.

As with the case of CIT, evidence showed that using metaheuristic methods with the generation process lead to efficient strategies~\cite{AFZAL2009957}. However, those strategies may suffer from the tuning of search parameters. More recently, we have developed a strategy for CIT that overcome this shortage by using an adaptive fuzzy interface system to tune the searching parameters \cite{Mahmoud2015}. Although useful, the strategy may not suit the generation of efficient VSCA due to the change of complexity during the generation and searching process when the interaction strength changes its value. Hence, there is a need for designing a new strategy that takes into account these difficulties with VSCA.

Bestoun et al.~\cite{Ahmed2011} have used PSO for VS interaction test suite generation. Their strategy is composed of two main algorithms, namely the ``Interaction Elements Generation Algorithm'' and the ``Test Suite Generation Algorithm''. The Interaction Elements Generation Algorithm computes and stores the interaction elements to be used for the test coverage evaluation. The Test Suite Generation Algorithm customizes the PSO algorithm for test suite generation. The various parameters of the PSO algorithm are set according to best practices recommended by other studies. Wu et al.~\cite{Wu2015} extended the set-based PSO for covering array generation. They named it as discrete PSO (DPSO). In DPSO, a candidate test case is represented by a particle's position and the particle's velocity is changed to a set of t-way schemas with probabilities. In addition, the authors have introduced two auxiliary strategies (particle reninitialization and additional evaluation of \textit{gbest}) to improve PSO performance. They compared DPSO with a conventional PSO (CPSO). A CPSO is designed to find optimal or near-optimal solutions in a continuous space. DPSO was found to produce smaller covering arrays than CPSO and other evolutionary algorithms.

\section{PSO Performance monitoring} \label{PSOPerformanceAnalysisSection}

The speed of finding an optimum solution in PSO is mainly dependent on the values of learning factors ($c_1 , c_2$), momentum weight factor ($w$) and location of the tested candidates in the search space. As a metaheuristic method, selecting its right learning and momentum weight factor values is the challenge of predicting the search direction that is assigned by the guided stochastic behavior. In Metaheuristics, the term ``exploration'' represents the method\textquoteright s ability to locate global values in the search space. On the other hand, the term ``exploitation'' represents methods' ability to locate local values in the search space \cite{Wang2014}. Furthermore, after analyzing the PSO performance under a range of learning and momentum weight factor values,  it can be concluded that PSO's exploration and exploitation capability is strongly affected by the learning and momentum weight factor values. By explicitly detailing the heuristic behavior of PSO, it is observed that momentum weight ($w$) is the exploration affecting element of the search by controlling the velocity of the particles, whereas the learning factors ($c_1, c_2$) are the exploitation elements \cite{Clerc2002}. For the sake of comparison, testing the convergence of PSO with regards to changing the value of $w$, a linearly has been witnesses to divert the search direction from the local to the global domain by \cite{Raaphorst2013}. In fact, $w$ has the most influential effect on the quality of the final results. Evidence showed that having a robust tuning of $w$ can lead to smoother deviation though the best solution \cite{Afshin2010,Mahmoud2015}. To this end, in most studies on PSO in the literature, the value of $c_1 and c_2$ are set to 2, and the value of $w$ is varied from 0.9 that help in the global search to 0.1 that assist in the local search.

In a nutshell, careful selection of $w$ is required to ensure that the global and local search are maintained balanced to reduce the convergence time. Although the route for reaching the optimum values for the objective is primarily drawn by selecting the right values of $w$, however depending on the problem in question and the size of it search space, variable yet accurate values of $w$, side by side with a proper swarm size, are required to enhance the performance of PSO.

\section{The Strategy} \label{StrategySection}

\subsection{Fuzzy Adaptive Swarm VS-CIT} \label{FuzzyAdaptive VS-CIT}

Based on the facts identified by analyzing the PSO's performance in \ref{PSOPerformanceAnalysisSection}, it is evident that variable weight momentum values can enhance the performance of PSO. This variation, in turn, needs to be guided and responsive to the search performance. Accordingly, the variables adaptation can be achieved by a Fuzzy rule-based system that can be built to correlate information about the PSO performance in making accurate variables adaptation. This can be built as part of an online performance monitoring mechanism to monitor and control the performance of PSO. Having such an intelligent monitoring system allows for extra control on the search performance, as well as avoidance of local minima problems. 

Fuzzy-based Self-Adaptive VS-CIT is an enhanced version of PSO and an extension to our previous work on CIT \cite{Mahmoud2015}. It is composed of an online performance monitoring mechanism to alter its momentum weight factor based on its performance and convergence measures. Studies such as \cite{Clerc2002,WangAndGeng2014} have revealed that small momentum weight values $w$ combined with large learning factors ($c_1$ and $c_2$) value direct the search towards local values. On the contrary, a significant value of the momentum weight $w$ combined with small learning factors ($c_1$ and $c_2$) value helps the optimization process to converge faster by focusing on the local domain of the identified optimum values. Ghanizadeh. \textit{el. al.} \cite{Afshin2010} suggested to vary the value of $w$ varied from 0.9 to 0.1 and make $c_1$ and $c_2$ to 2.

This fact concludes that applying these rules successfully on the PSO can control its performance by taking the right search path. It is hypothesized that this can overcome the local minimum and slow adaptation problems of PSO. These rules can be applied to achieve monitoring and control on PSO's performance by employing a Mamdani type Fuzzy Inference System (FIS) \cite{Camastra2015,MOHAMMAD201523} to generate control actions on $w$. 

The adaptation is applied concurrently to weight momentum factor via an individual FIS, to adapt $w$, based on the developer's experience. Each learning factor is changed at each PSO's adaptation interval. The proposed FISs are built to monitor the performance of the PSO and adjust $w$, to overcome the optimization process problems and improve its efficiency.

Therefore, the mechanism is designed to monitor the velocity update performance by counting the number of iterations that witnessed no change in fitness values. A normalization process was applied to the information about the number of constant fitness values in a velocity updates before introducing this information to the selected FIS as described in the below equation:

\begin{equation}
\centering
NorNUBF=\frac{NUBF_{Max}-NUBF_{k}}{NUBF_{k}}
\end{equation}

where NorNUBF is a monitoring index, which is implemented to read the  proportion of the number of iterations that have unchanged fitness values with respect to the total number of iterations proposed in the velocity updates loop. Accordingly, UBFk is the number of iterations that have unchanged best fitness values since the last fitness update, and NUBFmax is the maximum expected number of iterations in the velocity updates loop, which in turn represents the search space for finding global minimum (Gbest).

To simplify the proposed FIS to achieve the required adaptation, the normalization is implemented against the highest possible number of iterations that encountered no change in the fitness value, which is the value of the velocity iteration loop. Based on the literature \cite{Liang2006,Windisch2007,TSAI2017371,LYNN2017533}, we choose a proper range of operation of  $w$, such that the maximum inertia weight $w_{max}$ equal to $0.9$

\begin{center}
$0.1\le w\le$0.9
\par\end{center}

It is the value of inertia weight ($w_{adaptive}$) that is adaptively varied based on the maximum inertia, $w_{max}=0.9$. The distances $d_1$ and $d_2$ are based on Euclidian distance between the current vector \textbf{\textit{x}} with best particle $p_{best}$ and global best $g_{best}$ respectively. To this end, a normalization processes was applied on the information about the fitness functions as follows:

\begin{equation}
\centering
{\rm{Normalized}}\;{\rm{Current}}\;{\rm{Fitness(NCF)  = }}\frac{{{\rm{Current}}\;{\rm{Fitness  -  Min}}\;{\rm{Fitness}}}}{{{\rm{Max}}\;{\rm{Fitness  -  Min}}\;{\rm{Fitness}}}} \times 100
\end{equation}

In case of $w_{adaptive}$, by considering how it varies and what are its affecting variation factors, the change value required is determined based on the distance between the measured and targeted values $d_1$ and $d_2$ as in Equations \ref{D1Equation} and \ref{D2Equation}.

\begin{equation}
\label{D1Equation}
\centering
{d_1} = \frac{{\left| {{p_{best}} - x} \right|}}{{{\rm{Max}}\;{\rm{Distance}}}} \times 100
\end{equation}

\begin{equation}
\label{D2Equation}
\centering
{d_2} = \frac{{\left| {{g_{best}} - x} \right|}}{{{\rm{Max}}\;{\rm{Distance}}}} \times 100
\end{equation}

The range of $w_{adaptive}$ change can also vary depending on the objective function of the optimization problem in question. However, the general mathematical formulation of the $w_{adaptive}$ change is as in Equation \ref{WAdaptiveEquation}.

\begin{equation}
\label{WAdaptiveEquation}
\centering
{w_{adaptive}} = \frac{{{w_{selection}}}}{{{\rm{100}}}} \times {w_{\max }}
\end{equation}

\begin{figure}
\centering
\includegraphics [width=4.3 in]{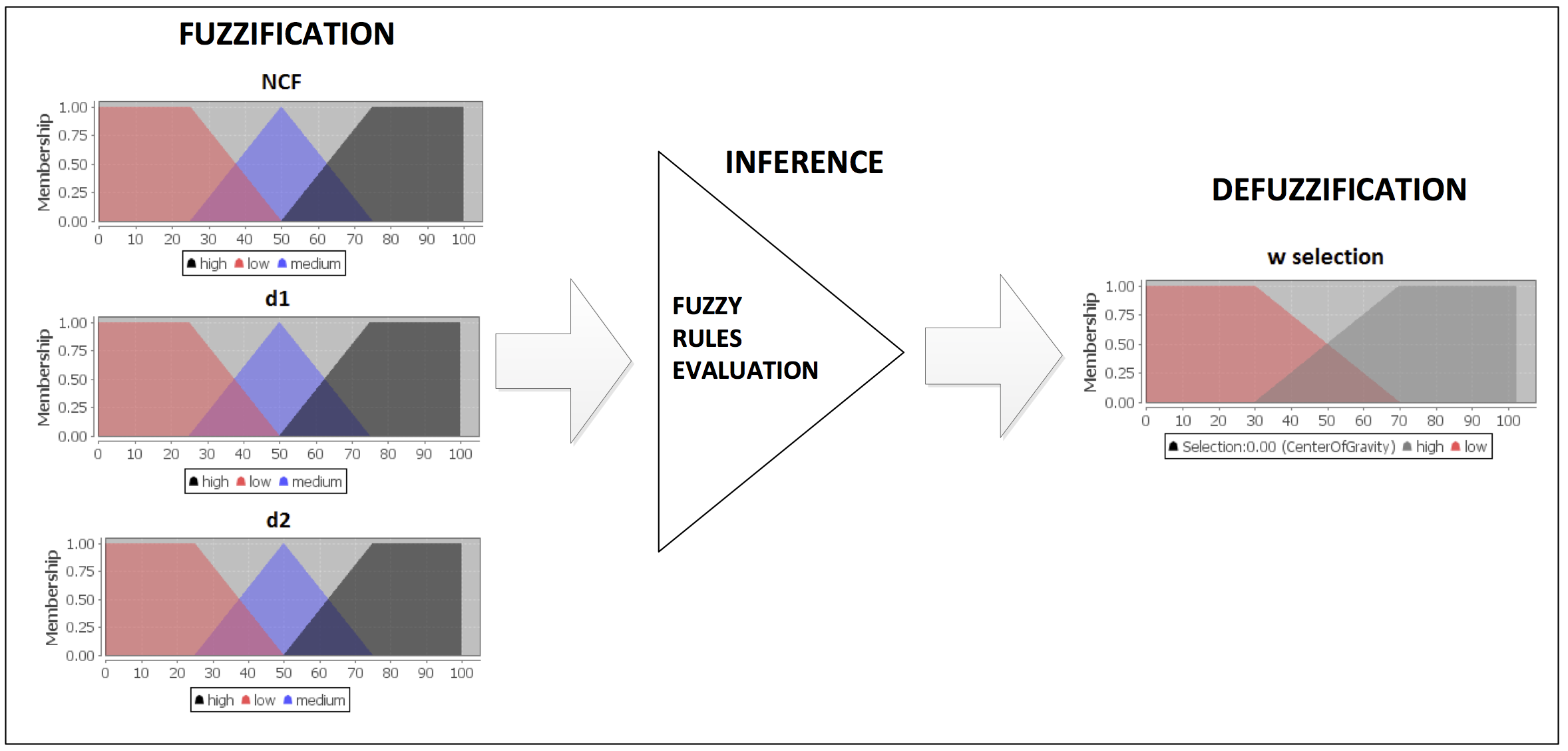}

\caption{The membership functions design of the two input one output $w$ adaptation FIS}
\label{Membership}
\end{figure}

Figure \ref{Membership} shows the membership functions of the $w_{adaptive}$
adaptation FIS.

Four defined fuzzy rules are exploited in the inference evaluation fuzzy rules based on the following scenarios:

\begin{itemize}
\item \textbf{Rule 1}: $NCF = low$, $d_1 = low$, $d_2 = low$.  The search is near convergence, local search is required. The value of $w = low$.
\item \textbf{Rule 2}: $NCF = not\;low$, $d_1 = low$, $d_2 = low$.  The search is trapped in local minima. The value of $w = high$.
\item \textbf{Rule 3}: $NCF = medium$, $d_1 = low$, $d_2 = not\;low$.  The search is exploring for new best. The value of $w = high$.
\item \textbf{Rule 4}: $NCF = high$, $d_1 = high$, $d_2 = high$.  The search is exploring for new best. The value of $w = high$.
\end{itemize}

Generally, this strategy divides the search space into a number of regions reflecting the number of particles in the swarm. The accuracy of these divisions will significantly depend on the complexity of the optimization problem. Initially finding the $p_{best}$ in the swarm is the first stage in the search process. Subsequently, velocity updates aim at finding the $g_{best}$ by creating a search direction driven by the weight momentum factor $w$.

The proposed FIS are called during the velocity updates to monitor the performance and adapt the weight momentum factor accordingly. The performance measures are translated as two input variables to the FIS to generates one output variable, which is the adaptation value.

The complete adaptation mechanism during the PSO search process is illustrated in Algorithm \ref{FuzzyPSOAlgorithmCode}.

\begin{algorithm}
\scriptsize
\label{FuzzyPSOAlgorithmCode}
 \caption{Pseudo code for Fuzzy PSO}

 \KwIn{Interaction strength ($t$), parameter ($k$) and its corresponding value ($v$)}
 \KwOut{Final test suite $F_s$}
 
 Initialize the population of the required $t-tuples$, $I = \left\{ {{I_1},{I_2}, \ldots ,{I_M}} \right\}$
 
 Initialize $\Theta_{max} $ iteration, and population size $S$
 
 Initialize the random population of solutions, $X = \left\{ {{X_1},{X_2}, \ldots ,{X_S}} \right\}$
 
 Initialize the max inertial weight $w_{max}$, and the learning factors $c_1$, $c_2$

 Initialize the population velocity $V = \left\{ {{V_1},{V_2}, \ldots ,{V_S}} \right\}$
 
 Choose global $X_{gbest}$ from the population $X = \left\{ {{X_1},{X_2}, \ldots ,{X_S}} \right\}$ 
 
 Set local best $X_{pbest} = X_{gbest}$
 
 Define Fuzzy Rules
 
 Compute max Euclidean distance based on the defined parameter $k$
 
  \While {all interaction tuples ($I$) are not covered}{
  
 \For {iteration =1 to $\Theta_{max} $}{
 
 \For {$i=1$ to $S$}{

NCF =Compute percentage NCF

$d_1$=Compute percentage  Euclidean distance  between $X_i$  and $X_{pbest}$

$d_2$= Compute percentage Euclidean distance  between $X_i$ and $X_{gbest}$ 

Fuzzify inputs NCF, $d_1$ and $d_2$ based on the defined membership functions

Apply fuzzy rules

Defuzzify to produce crisp output for percentage $w_{selection}$

Set $w=w_{selection} * w_{max} /100$

Update the current velocity according to $V_i^{(t+1)} = w \cdot V_i^{(t)} + c_1 \cdot rand(0,1) \cdot (X_{lbest}^{(t)} - X_i^{(t)}) + c_2 \cdot rand(0,1) \cdot (X_{gbest}-X_i^{(t)})$

Update the current population according to $X_i^{(t+1)} = X_i^{(t)} + V_i^{(t+1)}$

 \If {$f(X_i^{(t+1)}) > f(X_i^{(t)})$}{
 $X_i^{(t)} = X_i^{(t+1)}$
 } 
 
 \tcc{ ****** Update Global Best******** }
 
 \If {$f(X_i^{(t)}) > f(X_{gbest})$}{
$X_{gbest} = X_i^{(t)}$ 
 }

 }
 
Add $X_{gbest}$ in the final test suite $F_s$

 }
  
  }

\end{algorithm}

For the sake of testing the performance our strategy, it was applied to the CA construction, which represents a useful application of combinatorial optimization. CA is a mathematical model that has been used successfully in different real applications such as testing configurable software systems \cite{4564473}, controlling DC servo motors \cite{sahib2014application},  gene expression regulation \cite{shasha2001using}, advance material testing \cite{ziegel2003experimental}, and many other applications. By formulating this problem mathematically and implementing the strategy, the search process is detailed as follows:

\begin{itemize}
\item Initially, the $t-tuple$ list is generated base on the number of input parameters and values for each parameter. The detail of $t-tuple$ list generation is been given in Section \ref{PairGenerationSection}. At this stage, the $t-tuple$, list and then the random solutions generated by the swarm are tested against the number of rows it covers in the $t-tuple$. This process is repeated for a defined number of iterations until maximum coverage is achieved on all rows. This in turn represents $l_{best}$. 
\item Further test is conducted during the velocity update process to find the $g_{best}$. 
\item Throughout the velocity updates process, FIS are called to present the knowledge of the developer about the correlation between the performance measures to provide the required updates for $w$. 
\item As soon as a coverage is achieved, all covered rows are identified and deleted from the $t-tuple$ to facilitate quick search in the subsequent iteration. This is mainly to reduce the computation requirements and increase the accuracy of the process. The deletion, however, is only implemented if the solution is confirmed to be the optimum at the end of the PSO search process, including the velocity updates. 
\item As the covered rows deleted, they are also indexed as not available for subsequent covering trials. In this way, the covering tests only consider the remaining rows in the $t-tuples$. The covering computational time will be decreased exponentially along the covering process, and hence speed up the covering process.
\end{itemize}
%%%%%%%%%%%%%%%%%%%%%%%%%%%%%%%%%%%%%%%%%%%%%%%%

\subsection{The Pair Generation Algorithm }\label{PairGenerationSection}

The pair generation algorithm is an essential algorithm in any VS-CIT strategy to generate the $t-tuples$. Few algorithms developed in the literature for pair generation. However, these were not efficient when the number of input parameters grows due to the long generation time. More recently, we have developed an efficient algorithms algorithm using a new stack and hash table data structure \cite{Bestoun2016Hash}. Here should be noted that the same algorithm could be used for the VS-CIT. The only different here is that the algorithm will be repeated for each combination strength starting from lower to higher strengths.

The algorithm starts by generating the input combinations. Then, it puts the corresponding values for each combination respectively. The input combinations are generated base on the interaction strength $t$. Here, the algorithm takes $k$ input parameters of the SUT and the wanted interaction strength to generate the $t$-combinations of them. Algorithm \ref{ParameterCombinationAlgorithm} shows the steps of this algorithm in detail.

\begin{algorithm} [H]  \label{ParameterCombinationAlgorithm} 
\scriptsize
\linespread{1.1}\selectfont

 \KwIn{Input-parameters $k$ and combination strength $t$}
 \KwOut{All $t$-combinations of $k$ where $k={k_1 , k_2 , k_3 , ..., k_n}$ }
 Let Comb be an array of length $t$\;
 Let $i$ be the index of Comb array\;
 Create a stack $S$\;
 $S \leftarrow 0$\;
 \While {$S\neq null$}{
 
$i$ =(the length of $S -1$)\;
$v$ = pop the stack value\;

\While {pop value $< k$}{
set Comb of index ($i$) to $v$\;
$i\leftarrow i + 1$\;
$ v\leftarrow v + 1$\;
push $v$ to stack\;

\If{$i=t$}{
Add Comb to final array\;
break\;
}

}
}
 \caption{Parameter Combination Generator}
\end{algorithm}

As shown in Algorithm \ref{ParameterCombinationAlgorithm}, the stack data structure is used with the algorithm to avoid the out of memory problem when the number of parameters grows. The stack used here as a temporary memory to hold the parameters for few iterations. The parameters ``pushed'' into the stack and then ``pop'' when the algorithm iterate. As showed in (Steps $1-2$) in the algorithm, an array is created with length ``i'' for indexing during the iteration. The stack itself is created in (Step $3-4$) then the first parameter is pushed inside it. The index number ``i'' of the Comb array is set to length of $S − 1$ and the value ``v'' of this index ``i'' was set to the top value in the stack (i.e. pop) until ``v'' is less than ``k'' (Steps $6–9$). The algorithm will iterate until the stack will be empty (Step 5) and it will increment the value of ``i'' and ``v'' with the iteration.

\begin{figure}
\centering
\includegraphics [width=4 in]{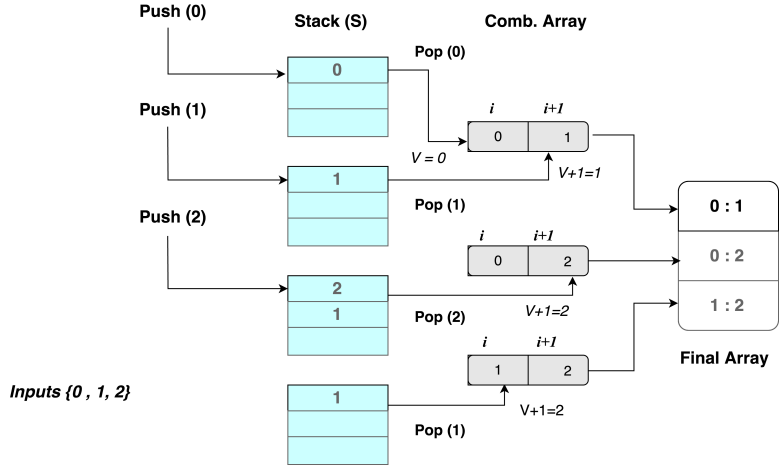}

\caption{A running example for Algorithm \ref{ParameterCombinationAlgorithm}}
\label{PairAlgorithmGraph}
\end{figure}

Figure \ref{PairAlgorithmGraph} shows a running example with three inputs {0, 1, 2} to illustrate the work of the algorithm clearly. Clearly, we can see the fist parameter (0) pushed into the stack. The stack then starts to pop its last value into the ``i+1'' index of Comb array. Then, by incrementing the iteration, the ``v+1'' value is pushed onto the stack. The algorithm is continuing this process until the stack gets empty. In this example, the final Comb array will contain the 2-combinations of the input parameters $\lbrace(0:1) , (0:2) , (1:2)\rbrace$.

\begin{figure}
\centering
\includegraphics [width=2 in]{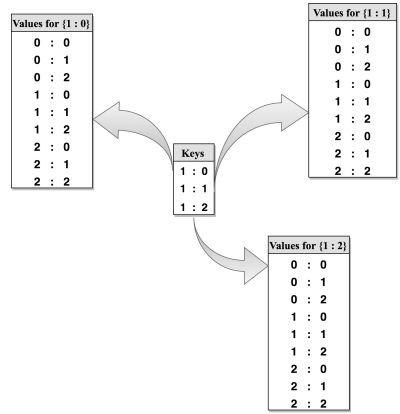}

%\centerline{\fbox{\hbox to 20pc{\vbox to 10pc{}}}}
\caption{The $<Key, value>$ pair structure in the hash table for the example in Figure \ref{PairAlgorithmGraph}}
\label{PairAlgorithmExample}
\end{figure}

Hash table is used to store the t-tuples list to facilitate and speed up the searching process. Here, the search process is done with the $<Key, value>$ pair. Each element in the Comb array is stored as a key value in the hash table. The tuple list for each corresponding key is stored array list in the value of the corresponding key of the hash table. To illustrate this design of the data structure, Figure \ref{PairAlgorithmExample} shows the $<Key, value>$ pair in the hash table for the example in Figure \ref{PairAlgorithmGraph}. We assume that each input parameter has three level of configurations $(0, 1, and 2)$.

%%%%%%%%%%%%%%%%%%%%%%%%%%%%%%%%%%%%%%%%

\section{Empirical Evaluation} \label{Evaluation}

We opt to benchmark our strategy (namely FPSO) against existing PSO based strategies. Those versions of PSO are, PSTG \cite{Ahmed2011,ahmed2010pstg} manual tuning of the parameter , Conventional PSO (CPSO)   and Discrete PSO (DPSO)  \cite{Wu2015}) using the variable strength configuration.  We opt for the variable strength configuration as published in Wu et al \cite{Wu2015}.

Our platform for experiment comprises of a PC running Windows 10, CPU 2.9 GHz Intel Core i5, 16 GB DDR3, 1867 MHz  RAM and a 512 MB of flash HDD.  In the interest of fairness, we do not consider the execution time owing to differences in the running environment, data structure, and language implementation. Unlike the time comparison, we believe the size comparison is fair since the generated size does not depend on and influence by the implementation and the deployed running system. In fact, most authors would have reported their best size results for comparison. In line with the published results and the evidence in the literature, we have run each experiment 30 times (i.e. the same number of iteration with Wu et al \cite{Wu2015}) and report the best (as bold cells) and the mean size (as shaded cells) to give a better indication of the performance of the strategies of interest. Tables \ref{ResultsTable1} through \ref{ResultsTable3} highlights our experimental results.

\begin{table}
\centering
\scriptsize 
\caption{Comparison with other PSOs for VCA (N; 2, $3^{15}$, \{C\}) }
\label{ResultsTable1}
\begin{tabular}{|c||c|c||c|c||c|c||c|c|}
\hline 
C & \multicolumn{2}{c||}{FPSO} & \multicolumn{2}{c||}{PSTG \cite{Ahmed2011}} & \multicolumn{2}{c||}{CPSO \cite{Wu2015}} & \multicolumn{2}{c|}{DPSO \cite{Wu2015}}\tabularnewline
\hline 
\hline 
 & Best & Mean & Best & Mean & Best & Mean & Best & Mean\tabularnewline
\hline 
$\phi$ & \textbf{18}  & 19.40  & 19  & 20.92  & 19  & 20.07  & \textbf{18}  & \cellcolor{blue!5}18.63 \tabularnewline
\hline 
CA (3, $3^{3}$) & \textbf{27} & 27.39 & \textbf{27} & 27.50 & 27 & 27.47  & 27 & \cellcolor{blue!5}27.27\tabularnewline
\hline 
\multirow{1}{*}{CA $(3,3^{3}){}^{2}$} & \textbf{27} & \cellcolor{blue!5}27.73  & \textbf{27} & 27.94 & \textbf{27} & 27.93 & \textbf{27} & 27.83\tabularnewline
\hline 
CA $(3,3^{3}){}^{3}$ & \textbf{27} & 28.05 & \textit{27} & 28.13  & \textbf{27} & 28.13  & \textbf{27} & \cellcolor{blue!5}28.00\tabularnewline
\hline 
CA (3, $3^{4}$) & 30 & 31.52 & 30 & 31.47 & 29 & 33.13 & \textbf{27} & \cellcolor{blue!5}31.43\tabularnewline
\hline 
CA (3, $3^{5}$) & 39 & 41.11 & \textbf{38} & \cellcolor{blue!5}39.83  & 39 & 41.23 & \textbf{38} & 40.93\tabularnewline
\hline 
CA (3, $3^{6}$) & \textbf{43} & \cellcolor{blue!5}45.00 & 45 & 46.42 & 44 & 46.27 & \textbf{43} & 45.70\tabularnewline
\hline 
CA (3, $3^{7}$) & 48 & 50.61 & 49 & 51.68 & 49 & 50.97 & \textbf{47} & \cellcolor{blue!5}49.87\tabularnewline
\hline 
CA (3, $3^{9}$) & \textbf{56} & \cellcolor{blue!5}58.00 & 57 & 59.11 & 58 & 60.36 & \textbf{56} & 58.18\tabularnewline
\hline 
CA (4, $3^{4}$) & \textbf{81} & \cellcolor{blue!5}81.03 & \textbf{81} & 82.21 & \textbf{81} & 81.07 & \textbf{81} & \cellcolor{blue!5}81.03\tabularnewline
\hline 
CA (4, $3^{5}$) & 94 & 99.82 & 97 & 99.31 & 97 & 101.03 & \textbf{85} & \cellcolor{blue!5}94.50\tabularnewline
\hline 
CA (4, $3^{7}$) & \textbf{152} & 156.87 & 158 & 160.31 & 154 & 157.93 & \textbf{152} & \cellcolor{blue!5}156.83 \tabularnewline
\hline 
\end{tabular}

\end{table}

\begin{table}
\centering
\scriptsize 
\caption{Comparison with other PSOs for VCA (N; 3, $3^{15}$, \{C\}) }
\label{ResultsTable2}
\begin{tabular}{|c||c|c||c|c||c|c||c|c|}
\hline 
C & \multicolumn{2}{c||}{FPSO} & \multicolumn{2}{c||}{PSTG \cite{Ahmed2011}} & \multicolumn{2}{c||}{CPSO \cite{Wu2015}} & \multicolumn{2}{c|}{DPSO \cite{Wu2015} }\tabularnewline
\hline 
\hline 
 & Best & Mean & Best & Mean & Best & Mean & Best & Mean\tabularnewline
\hline 
$\phi$ & \textbf{72}  & 75.69  & 75  & 78.69  & 75  & 76.73  & \textbf{72}  & \cellcolor{blue!5}73.97 \tabularnewline
\hline 
CA (4, $3^{4}$) & \textbf{86} & 91.55 & 91 & 91.80 & \textbf{86} & 91.45 & \textbf{86} & \cellcolor{blue!5}89.83\tabularnewline
\hline 
\multirow{1}{*}{CA $(4,3^{4}){}^{2}$} & \textbf{88} & \cellcolor{blue!5}90.60 & 91 & 92.21 & 90 & 92.93 & \textbf{88} & 90.77\tabularnewline
\hline 
CA (4, $3^{5}$) & 109 & 113.90 & 114 & 117.30 & 108 & 114.03 & \textbf{107} & \cellcolor{blue!5}111.17\tabularnewline
\hline 
CA (4, $3^{7}$) & 155 & 159.30 & 159 & 162.23 & 156 & 159.40 & \textbf{152} & \cellcolor{blue!5}158.57\tabularnewline
\hline 
CA (4, $3^{9}$) & \textbf{193} & 196.20 & 195 & 199.28 & \textbf{193} & 196.57 & \textbf{193} & \cellcolor{blue!5}196.00\tabularnewline
\hline 
CA (4, $3^{11}$) & \textbf{225} & \cellcolor{blue!5}227.46 & 226 & 230.64 & 227 & 229.3 & \textbf{225} & 227.50\tabularnewline
\hline 
\end{tabular}

\end{table}

\begin{table}
\centering
\scriptsize 
\caption{Comparison with other PSOs for VCA (N; 2, $4^{3}$ $5^{3}$ $6^{2}$,
\{C\})}
\label{ResultsTable3}
\begin{tabular}{|>{\raggedright}m{2.5cm}||c|c||c|c||c|c||c|c|}
\hline 
C & \multicolumn{2}{c||}{FPSO} & \multicolumn{2}{c||}{PSTG \cite{Ahmed2011}} & \multicolumn{2}{c||}{CPSO \cite{Wu2015}} & \multicolumn{2}{c|}{DPSO \cite{Wu2015}}\tabularnewline
\hline 
\hline 
 & Best & Mean & Best & Mean & Best & Mean & Best & Mean\tabularnewline
\hline 
$\phi$ & \textbf{40} & \cellcolor{blue!5}42.28  & 42  & 43.60  & 41  & 43.30  & \textbf{40}  & 42.30 \tabularnewline
\hline 
CA (3, $4^{3}$)  & \textbf{64} & \cellcolor{blue!5}64.00 & \textbf{64} & 65.50 & \textbf{64} & 64.30 & \textbf{64} & 64.00\tabularnewline
\hline 
\multirow{1}{2.5cm}{CA (3, $4^{3}$$5^{2}$)} & \textbf{117} & \cellcolor{blue!5}124.30 & 124 & 126.60 & 123 & 127.40 & 119 & 124.70\tabularnewline
\hline 
CA (3, $4^{3}$), 

CA (3, $5^{3}$) & \textbf{125} & \cellcolor{blue!5}125.00 & \textbf{125} & 127.90 & \textbf{125} & 125.10 & \textbf{125} & \cellcolor{blue!5}125.00\tabularnewline
\hline 
CA (3, $4^{3}$ $5^{3}$ $6^{1}$)  & 206 & 210.00 & 206 & 210.20 & 207 & 212.70 & \textbf{203} & \cellcolor{blue!5}207.50 \tabularnewline
\hline 
CA (3, $4^{3}$), 

CA (4, $5^{3}$ $6^{1}$), & \textbf{750} & 750.26 & \textbf{750} & 755.70 & \textbf{750} & \cellcolor{blue!5}750.10 & \textbf{750} & 750.80\tabularnewline
\hline 
CA (4, $4^{3}$ $5^{2}$)  & 462 & \cellcolor{blue!5}465.20 & 472 & 478.10 & 467 & 475.40  & \textbf{440} & 450.60\tabularnewline
\hline 
\end{tabular}

\end{table}

\begin{figure}
\centering
\includegraphics [width=3 in]{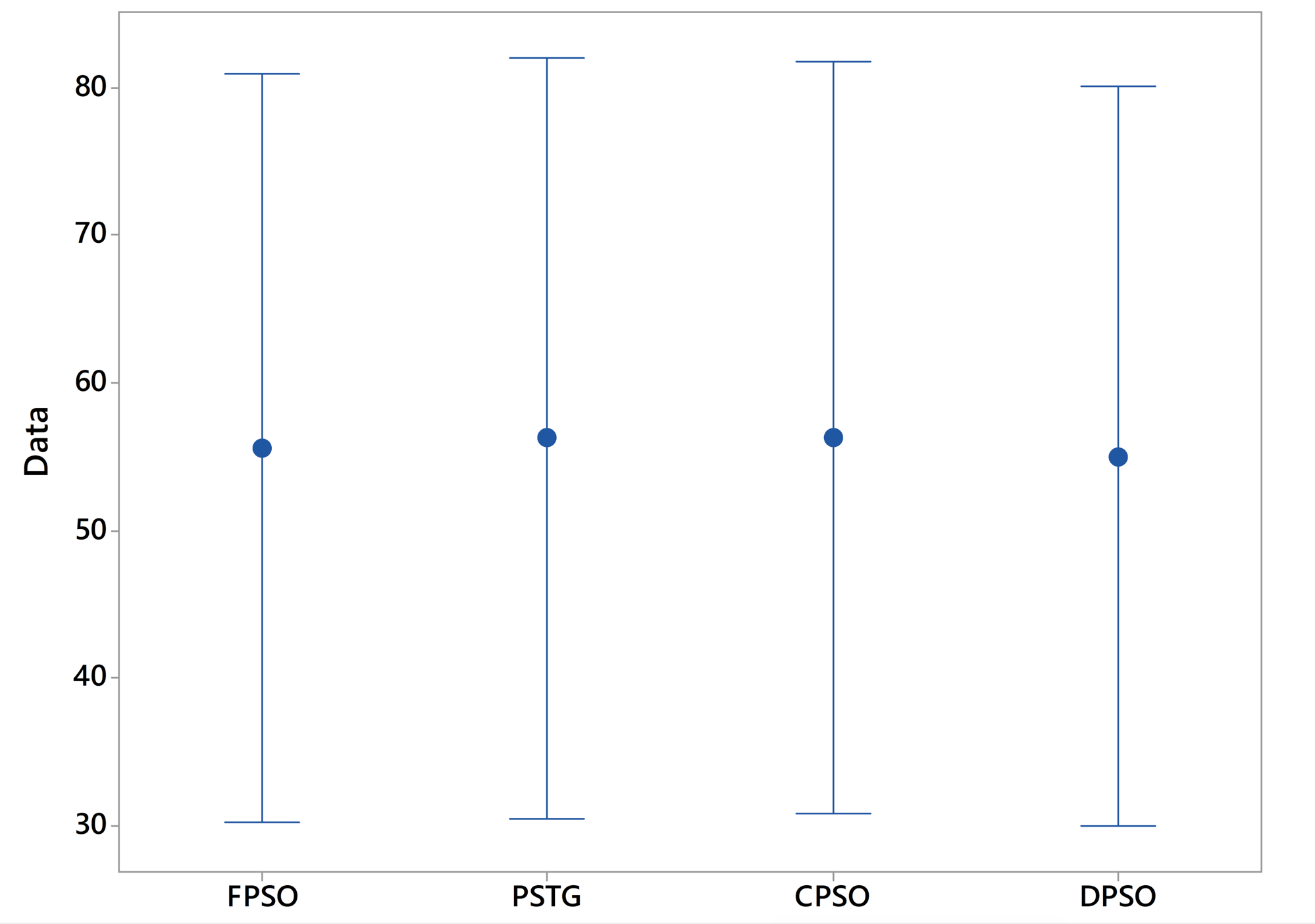}

\caption{Interval plot for Table \ref{ResultsTable1}}
\label{Interval_Plot_1}
\end{figure}

\begin{figure}
\centering
\includegraphics [width=3 in]{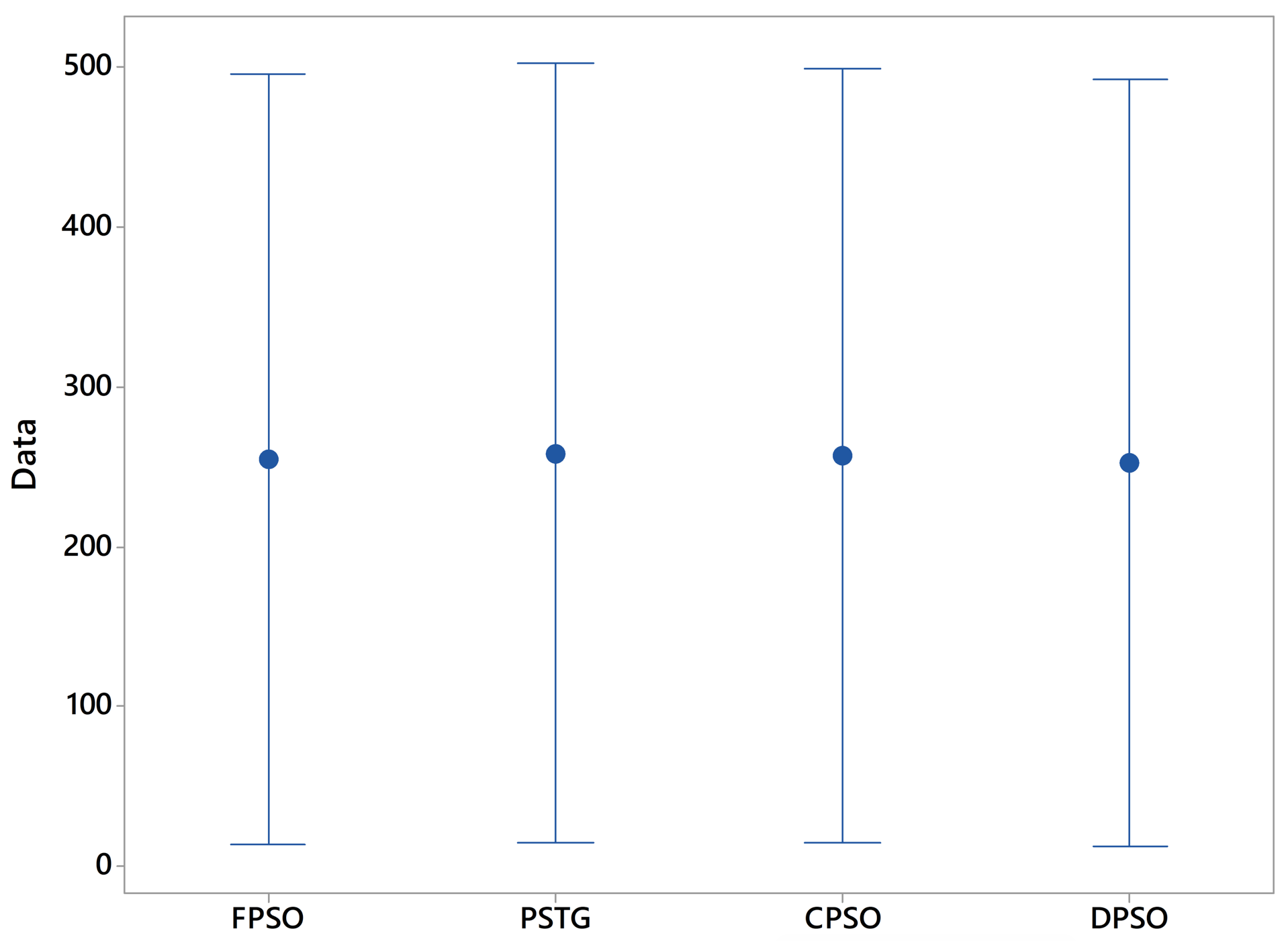}

%\centerline{\fbox{\hbox to 20pc{\vbox to 10pc{}}}}
\caption{Interval plot for Table \ref{ResultsTable2}}
\label{Interval_Plot_2}
\end{figure}

\begin{figure}
\centering
\includegraphics [width=3 in]{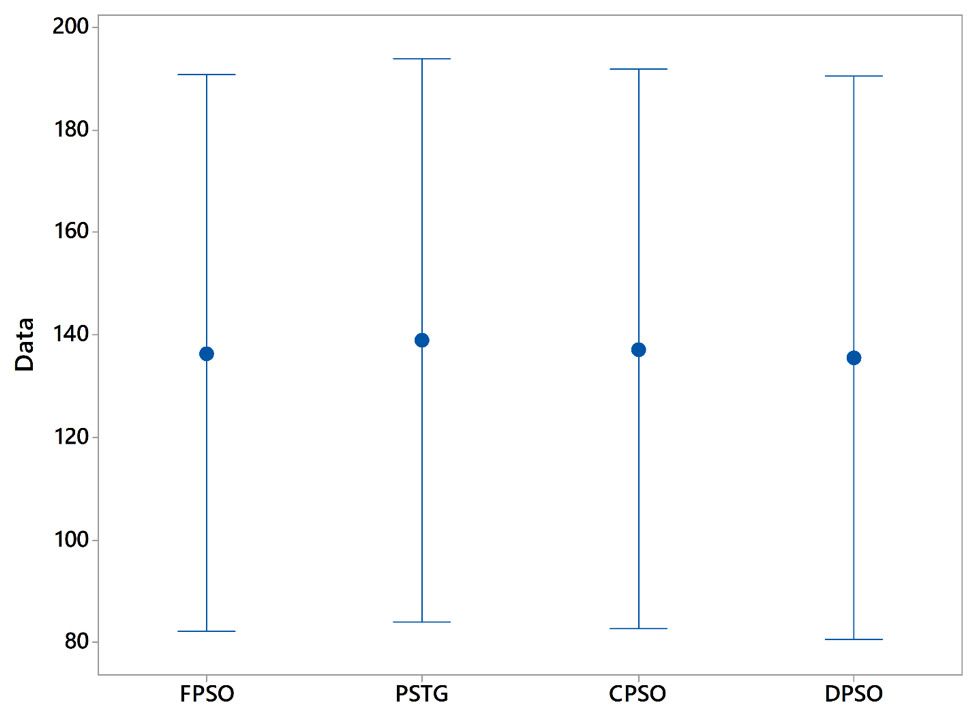}

%\centerline{\fbox{\hbox to 20pc{\vbox to 10pc{}}}}
\caption{Interval plot for Table \ref{ResultsTable3}}
\label{Interval_Plot_3}
\end{figure}

\section{Observation} \label{Discussion}

The following general observations, independent of the implementation and coding used, were made through conducting the experiments.

First, concerning best test size, DPSO gives the best overall results in all tables (refer to bold cells). Specifically, DPSO obtains 6 of the best overall size results for $VCA (N; 2, 3^{15}, CA (3, 3^7))$ and $VCA (N; 2, 3^{15}, CA (4, 3^5))$ in Table \ref{ResultsTable1}, $VCA (N; 3, 3^{15},$ $ CA (4, 3^5))$ and $VCA (N; 3, 3^{15}, CA (4, 3^7))$ in Table \ref{ResultsTable3}, $VCA (N; 2, 4^3 5^3  6^2,CA (3, 4^3 $  $5^3 6^1))$ and $VCA (N; 2, 43 53 62, CA (4, 43 52))$ in Table \ref{ResultsTable3}. Except for entries for $VCA (N; 2, 4^3 5^3$  $6^2,  CA (3, 4^3 5^2))$ in Table \ref{ResultsTable3} where FPSO gives the best overall result, DPSO also matches with all other best results in most other configurations. Considering the number of bold cells where the test sizes match with the best size reported by other strategies, FPSO comes in as the runner-up followed by CPSO and DPSO.

Second, concerning mean test size, a similar pattern can be observed (refer to highlighted cells). In particular, DPSO gives the overall best mean for most cases. In Table \ref{ResultsTable1}, DPSO obtains the best mean test sizes for 8 out of 12 entries (with 66.66 percent). Concerning Table \ref{ResultsTable2}, DPSO obtains the best mean test sizes for 5 out of 7 entries (with  71.42 percent). Interestingly, FPSO outperforms DPSO in terms of obtaining the best mean test sizes in Table \ref{ResultsTable3} with %5 
3 out of 7 entries %(with 71.42 percent) although there are 
(with 3 entries having the same mean).

Another subtle observation is the fact that the mean for PSTG and CPSO have always been inferior to that of DPSO and FPSO with the exception of $VCA (N; 2$  $, 3^{15}$, $CA (3, 3^5))$ and $VCA (N; 2, 4^3 5^3 6^2$, $CA (3, 43), $ $CA (4, 5^3 6^1))$.

From the interval plots shown in Figure \ref{Interval_Plot_1} for Table \ref{ResultsTable1}, it is noted that DPSO manages to achieve the minimum overall average. The best performance was achieved based on the mean distribution at 95\% confidence levels. The FPSO comes in the second ranking while CPSO is in the third rank. Finally, PSTG show the worst overall mean distribution, hence, having the poorest size performance.

Based on the interval plots in Figure \ref{Interval_Plot_2} for Table \ref{ResultsTable2}, it shows that DPSO gives the minimum overall mean distribution at 95\%  confidence levels. Hence, the best performance was achieved. The FPSO get the second  ranking while CPSO and PSTG show similar performance with the worst overall mean distribution for this configurations.

From the mean distribution shown in the interval plot in Figure \ref{Interval_Plot_3} for Table \ref{ResultsTable3}, the DPSO is shown to give the minimum mean distribution at 95\%  confidence levels. The FPSO followed in the second position, then, CPSO in the third ranking. Finally, PSTG shows the poorest overall performance in terms of average size performance.

Summing up and reflecting on the work undertaken, we note that achieving performance for general PSO is not without costs. To be specific, DPSO appears to be superior to other PSO variants concerning test size performance. However, unlike other PSO variants, DPSO is a heavyweight strategy with 3 new controls (pro1 = parameter for velocity control, pro2 = parameter for fixed schema test selection and pro3 = parameter for invoking mutation) in addition to the generic parameters (social $c_1$, $c_2$ and inertia $w$). These parameters requires extensive tuning before good quality solution can be obtained. As such, the generality of DPSO can be an issue. In contrast , our developed FPSO is lightweight and does not require specific tuning. In fact, the $c_1$ and $c_2$ are fixed,  only the inertial weight ($w$) that is adaptively varied according to the need of the current search process (i.e. either exploring or exploiting  based on the defined fuzzy rules). In this manner, our approach is more generic and readily applicable to other optimization problems.

%%%%%%%%%%%%%%%%%%%%%%%%%%%%%%%%%%%%%%%%%%%%%%%%%%%
\section{Conclusions} \label{Conclusion}

This paper highlights the design and development of an adaptive fuzzy based PSO. The strategy represents a new approach to use adaptive mechanism for controlling exploration and exploitation of the PSO search based on the current landscape of search. Unlike existing fuzzy based approaches (e.g. \cite{Mahmoud2015}), our work only adaptively adjusts the inertial weight (w), while keeping the learning factor c1 and c2 constants. For this reason, our approach is more intuitive as compared to existing approaches (as our fuzzy rules can be expressed as straightforward linguistic constraints that are easy to understand and maintain). In fact, this approach could be applicable in many test generation processes in software engineering. The approach can also open a new direction for test generation in the context of search-based software engineering (SBSE).

Given the results mentioned above elaborated in the earlier section, we believe that further empirical evaluations of the fuzzy parameter tuning are warranted. It should be noted that the randomness of the search operator of all PSO variants can be an issue here. The best test size results can potentially be obtained by chance and only once out of many runs.  For this reason, we believe comparison amongst the mean could also be useful to shed light on the performance of our developed FPSO.

Apart from modifying how the original PSO performs the position updates through probabilistic means, the performance of DPSO can be attributed to its adoption of average Hamming distance similarity measure for deciding the best candidate test case t to be added the final test suite TS. If a test candidate t1 and t2 have the same weights, DPSO will select the candidate with the most different Hamming distance as the best test case. In this manner, DPSO potentially increases the chance of getting a more diversified solution to be generated.

CPSO and DPSO, given the pure PSO implementation, appear to suffer from poor convergence despite having been tuned extensively. For this reason, although they can match some of the best results, most of the results from CPSO and PSO are more inferior to that of DPSO and FPSO.

As the scope of future work, we are looking into three possibilities. Firstly, we are trying to investigate Sugeno fuzzy inference as an alternative to our Mamdani one. Furthermore, we are also interested in incorporating Hamming distance measure as part of our FPSO implementation similar to DPSO. Finally, we are also looking into adopting our FPSO for generating test suites for high constraints system involving software product lines.

\bibliography{bestounbib}

\begin{thebibliography}{10}
\expandafter\ifx\csname url\endcsname\relax
  \def\url#1{\texttt{#1}}\fi
\expandafter\ifx\csname urlprefix\endcsname\relax\def\urlprefix{URL }\fi
\expandafter\ifx\csname href\endcsname\relax
  \def\href#1#2{#2} \def\path#1{#1}\fi

\bibitem{BestounKamalFuzzy2017}
K.~Z. Zamli, F.~Din, S.~Baharom, B.~S. Ahmed, Fuzzy adaptive teaching
  learning-based optimization strategy for the problem of generating mixed
  strength t-way test suites, Engineering Applications of Artificial
  Intelligence 59~(C) (2017) 35--50.

\bibitem{Camastra2015}
F.~Camastra, A.~Ciaramella, V.~Giovannelli, M.~Lener, V.~Rastelli, A.~Staiano,
  G.~Staiano, A.~Starace, A fuzzy decision system for genetically modified
  plant environmental risk assessment using mamdani inference, Expert Syst.
  Appl. 42~(3) (2015) 1710--1716.

\bibitem{TSAKIRIDIS2017257}
N.~L. Tsakiridis, J.~B. Theocharis, G.~C. Zalidis, Deco3rum: A differential
  evolution learning approach for generating compact mamdani fuzzy rule-based
  models, Expert Systems with Applications 83 (2017) 257 -- 272.

\bibitem{KHOSRAVANIAN2016280}
R.~Khosravanian, M.~Sabah, D.~A. Wood, A.~Shahryari, Weight on drill bit
  prediction models: Sugeno-type and mamdani-type fuzzy inference systems
  compared, Journal of Natural Gas Science and Engineering 36 (2016) 280 --
  297.

\bibitem{Nie2011}
C.~Nie, H.~Leung, A survey of combinatorial testing, ACM Computing Surveys
  43~(2) (2011) 11:1--11:29.

\bibitem{Yilmaz2014}
C.~Yilmaz, S.~Fouche, M.~Cohen, A.~A. Porter, G.~Demir\"{o}z, U.~Ko\c{c},
  Moving forward with combinatorial interaction testing, Computer 47~(2) (2014)
  37--45.

\bibitem{Hervieu2016}
A.~Hervieu, D.~Marijan, A.~Gotlieb, B.~Baudry, Practical minimization of
  pairwise-covering test configurations using constraint programming,
  Information and Software Technology 71~(C) (2016) 129--146.

\bibitem{Kacker2013}
R.~N. Kacker, D.~R. Kuhn, Y.~Lei, J.~F. Lawrence, Combinatorial testing for
  software: An adaptation of design of experiments, Measurement 46~(9) (2013)
  3745 -- 3752.

\bibitem{Kuhn2013}
D.~R. Kuhn, R.~N. Kacker, Y.~Lei, Introduction to Combinatorial Testing, 1st
  Edition, Chapman \& Hall/CRC, 2013.

\bibitem{Ahmed2011}
B.~S. Ahmed, K.~Z. Zamli, A variable strength interaction test suites
  generation strategy using particle swarm optimization, Journal of Systems and
  Software 84~(12) (2011) 2171--2185.

\bibitem{ZamliISPaper2017}
K.~Z. Zamli, F.~Din, G.~Kendall, B.~S. Ahmed, An experimental study of
  hyper-heuristic selection and acceptance mechanism for combinatorial t-way
  test suite generation, Information Sciences 399~(C) (2017) 121--153.

\bibitem{1245373}
M.~B. Cohen, P.~B. Gibbons, W.~B. Mugridge, C.~J. Colbourn, J.~S. Collofello, A
  variable strength interaction testing of components, in: Proceedings 27th
  Annual International Computer Software and Applications Conference. COMPAC
  2003, 2003.

\bibitem{huang2013prioritizing}
R.~Huang, J.~Chen, T.~Zhang, R.~Wang, Y.~Lu, Prioritizing variable-strength
  covering array, in: Computer Software and Applications Conference (COMPSAC),
  2013 IEEE 37th Annual, IEEE, 2013, pp. 502--511.

\bibitem{Ahmed2012}
B.~S. Ahmed, K.~Z. Zamli, C.~P. Lim, Application of particle swarm optimization
  to uniform and variable strength covering array construction, Applied Soft
  Compututing 12~(4) (2012) 1330--1347.

\bibitem{BestounConstraints2017}
B.~S. Ahmed, L.~M. Gambardella, W.~Afzal, K.~Z. Zamli, Handling constraints in
  combinatorial interaction testing in the presence of multi objective particle
  swarm and multithreading, Information and Software Technology 86~(C) (2017)
  20--36.

\bibitem{AhmedBestoun2016}
B.~S. Ahmed, Test case minimization approach using fault detection and
  combinatorial optimization techniques for configuration-aware structural
  testing, Engineering Science and Technology, an International Journal 19~(2)
  (2016) 737 -- 753.

\bibitem{Wang2014}
H.~Wang, Q.~Geng, Z.~Qiao, Parameter tuning of particle swarm optimization by
  using taguchi method and its application to motor design, in: 2014 4th IEEE
  International Conference on Information Science and Technology, 2014, pp.
  722--726.

\bibitem{Xu2013}
G.~Xu, An adaptive parameter tuning of particle swarm optimization algorithm,
  Applied Mathematics and Computation 219~(9) (2013) 4560 -- 4569.

\bibitem{Bestoun2012IJICIC}
B.~S. Ahmed, K.~Z. Zamli, C.~P. Lim, Constructing a t-way interaction test
  suite using the particle swarm optimization approach, International Journal
  of Innovative Computing, Information and Control 8~(1 A) (2012) 431--451.

\bibitem{Mahmoud2015}
T.~Mahmoud, B.~S. Ahmed, An efficient strategy for covering array construction
  with fuzzy logic-based adaptive swarm optimization for software testing use,
  Expert Systems with Applications 42~(22) (2015) 8753--8765.

\bibitem{Cohen2003}
M.~B. Cohen, P.~B. Gibbons, W.~B. Mugridge, C.~J. Colbourn, Constructing test
  suites for interaction testing, in: Proceedings of the 25th International
  Conference on Software Engineering, ICSE '03, IEEE Computer Society,
  Washington, DC, USA, 2003, pp. 38--48.

\bibitem{Colbourn2010}
C.~J. Colbourn, G.~K{\'e}ri, P.~P. Rivas~Soriano, J.~C. Schlage-Puchta,
  Covering and radius-covering arrays: Constructions and classification,
  Discrete Applied Mathematics 158~(11) (2010) 1158--1180.

\bibitem{Sloane1993}
N.~J.~A. Sloane, Covering arrays and intersecting codes, Journal of
  Combinatorial Designs 1~(1) (1993) 51--63.

\bibitem{Kari2004}
K.~J. Nurmela, Upper bounds for covering arrays by tabu search, Discrete
  Applied Mathematics 138~(1-2) (2004) 143--152.

\bibitem{Ahmed2015}
B.~S. Ahmed, T.~S. Abdulsamad, M.~Y. Potrus, Achievement of minimized
  combinatorial test suite for configuration-aware software functional testing
  using the cuckoo search algorithm, Information and Software Technolgy 66~(C)
  (2015) 13--29.

\bibitem{Hartman2005}
A.~Hartman, Software and Hardware Testing Using Combinatorial Covering Suites,
  Springer US, Boston, MA, 2005, pp. 237--266.

\bibitem{Raaphorst2013}
S.~Raaphorst, Variable strength covering arrays, Ph.D. thesis, School of
  Electrical Engineering and Computer Science Faculty of Engineering,
  University of Ottawa, Ottawa, Canada (2013).

\bibitem{Bestoun-plos}
B.~S. Ahmed, M.~A. Sahib, L.~M. Gambardella, W.~Afzal, K.~Z. Zamli, Optimum
  design of {PI$\lambda$D$\mu$} controller for an automatic voltage regulator
  system using combinatorial test design, PLOS ONE 11~(11) (2016) 1--20.

\bibitem{4601539}
Z.~Wang, B.~Xu, C.~Nie, Greedy heuristic algorithms to generate variable
  strength combinatorial test suite, in: 2008 The Eighth International
  Conference on Quality Software, 2008.

\bibitem{Bryce2007}
R.~C. Bryce, C.~J. Colbourn, One-test-at-a-time heuristic search for
  interaction test suites, in: Proceedings of the 9th Annual Conference on
  Genetic and Evolutionary Computation, GECCO '07, ACM, New York, NY, USA,
  2007, pp. 1082--1089.

\bibitem{Forbes2008}
M.~Forbes, J.~Lawrence, Y.~Lei, R.~N. Kacker, D.~R. Kuhn, Refining the
  in-parameter-order strategy for constructing covering arrays, Journal of
  Research of the National Institute of Standards and Technology 113~(5) (2008)
  287.

\bibitem{Lei2008}
Y.~Lei, R.~Kacker, D.~R. Kuhn, V.~Okun, J.~Lawrence, Ipog-ipog-d: Efficient
  test generation for multi-way combinatorial testing, Software Testing
  Verification and Reliability 18~(3) (2008) 125--148.

\bibitem{Chen2009}
X.~Chen, Q.~Gu, A.~Li, D.~Chen, Variable strength interaction testing with an
  ant colony system approach, in: 2009 16th Asia-Pacific Software Engineering
  Conference, 2009, pp. 160--167.

\bibitem{Czerwonka2008}
J.~Czerwonka, Pairwise testing in the real world: Practical extensions to
  test-case scenarios,
  https://msdn.microsoft.com/en-us/windows/hardware/drivers/taef/pict-data-source,
  online; accessed 19 February 2017 (2008).

\bibitem{AFZAL2009957}
W.~Afzal, R.~Torkar, R.~Feldt, A systematic review of search-based testing for
  non-functional system properties, Information and Software Technology 51~(6)
  (2009) 957 -- 976.

\bibitem{Wu2015}
H.~Wu, C.~Nie, F.~C. Kuo, H.~Leung, C.~J. Colbourn, A discrete particle swarm
  optimization for covering array generation, IEEE Transactions on Evolutionary
  Computation 19~(4) (2015) 575--591.

\bibitem{Clerc2002}
M.~Clerc, J.~Kennedy, The particle swarm - explosion, stability, and
  convergence in a multidimensional complex space, IEEE Transactions on
  Evolutionary Computation 6~(1) (2002) 58--73.

\bibitem{Afshin2010}
A.~Ghanizadeh, S.~Sinaie, A.~A. Abarghouei, S.~M. Shamsuddin, Notice of
  retraction a fuzzy-particle swarm optimization based algorithm for solving
  shortest path problem, in: 2010 2nd International Conference on Computer
  Engineering and Technology, Vol.~6, 2010, pp. V6--404--V6--408.

\bibitem{WangAndGeng2014}
H.~Wang, Q.~Geng, Z.~Qiao, Parameter tuning of particle swarm optimization by
  using taguchi method and its application to motor design, in: 2014 4th IEEE
  International Conference on Information Science and Technology, 2014, pp.
  722--726.

\bibitem{MOHAMMAD201523}
R.~Mohammad, A.~Mostafa, M.~Abbas, H.~M. Farouq, Prediction of representative
  deformation modulus of longwall panel roof rock strata using mamdani fuzzy
  system, International Journal of Mining Science and Technology 25~(1) (2015)
  23 -- 30.

\bibitem{Liang2006}
J.~J. Liang, A.~K. Qin, P.~N. Suganthan, S.~Baskar, Comprehensive learning
  particle swarm optimizer for global optimization of multimodal functions,
  IEEE Transactions on Evolutionary Computation 10~(3) (2006) 281--295.

\bibitem{Windisch2007}
A.~Windisch, S.~Wappler, J.~Wegener, Applying particle swarm optimization to
  software testing, in: Proceedings of GECCO 2007: Genetic and Evolutionary
  Computation Conference, 2007, pp. 1121--1128.

\bibitem{TSAI2017371}
H.-C. Tsai, Unified particle swarm delivers high efficiency to particle swarm
  optimization, Applied Soft Computing 55 (2017) 371 -- 383.

\bibitem{LYNN2017533}
N.~Lynn, P.~N. Suganthan, Ensemble particle swarm optimizer, Applied Soft
  Computing 55 (2017) 533 -- 548.

\bibitem{4564473}
M.~B. Cohen, M.~B. Dwyer, J.~Shi, Constructing interaction test suites for
  highly-configurable systems in the presence of constraints: A greedy
  approach, IEEE Transactions on Software Engineering 34~(5) (2008) 633--650.

\bibitem{sahib2014application}
M.~A. Sahib, B.~S. Ahmed, M.~Y. Potrus, Application of combinatorial
  interaction design for dc servomotor pid controller tuning, Journal of
  Control Science and Engineering 2014 (2014) 4.

\bibitem{shasha2001using}
D.~E. Shasha, A.~Y. Kouranov, L.~V. Lejay, M.~F. Chou, G.~M. Coruzzi, Using
  combinatorial design to study regulation by multiple input signals. a tool
  for parsimony in the post-genomics era, Plant Physiology 127~(4) (2001)
  1590--1594.

\bibitem{ziegel2003experimental}
E.~R. Ziegel, Experimental design for combinatorial and high throughput
  materials development (2003).

\bibitem{Bestoun2016Hash}
B.~S. Ahmed, L.~M. Gambardella, K.~Z. Zamli, A new approach to speed up
  combinatorial search strategies using stack and hash table, in: 2016 SAI
  Computing Conference (SAI), 2016, pp. 1217--1222.

\bibitem{ahmed2010pstg}
B.~S. Ahmed, K.~Z. Zamli, Pstg: a t-way strategy adopting particle swarm
  optimization, in: Mathematical/Analytical Modelling and Computer Simulation
  (AMS), 2010 Fourth Asia International Conference on, IEEE, 2010, pp. 1--5.

\end{thebibliography}
%\begin{thebibliography}{00}

% \end{thebibliography}

\end{document}